# Extreme narrow magnetic domain walls in U ferromagnets: The UCoGa case

Petr Opletal, Klára Uhlířová, Ivo Kalabis, Vladimír Sechovský and Jan Prokleška[*]

*Charles University, Faculty of Mathematics and Physics, Department of Condensed Matter Physics, Ke Karlovu 5, 121 16 Prague 2, Czech Republic*

**Abstract**

Surface magnetic domains of a UCoGa single crystal during magnetization/demagnetization processes in increasing/decreasing magnetic fields were investigated by means of magnetic-force-microscopy (MFM) images at low temperatures. The observed domain structure is typical for a ferromagnet with strong uniaxial anisotropy. The evolution of magnetic domains during cooling of the crystal below $T_\mathrm{C}$ has also been manifested by MFM images. Analysis of the available data reveals that the high uniaxial magnetocrystalline energy in combination with the relatively small ferromagnetic exchange interaction in UCoGa gives rise to the formation of very narrow domain walls formed by the pairs of the nearest U neighbor ions with antiparallel magnetic moments within the basal plane. Since the very high anisotropy energy is a common feature of the majority of the uniaxial U ferromagnets, analogous domain-wall properties are expected for all these materials.

[*]corresponding author: prokles@mag.mff.cuni.cz

## Introduction

Magnetocrystalline anisotropy (MA) is manifested by locking the macroscopic magnetic moment in a certain direction of a crystal, the easy magnetization axis (easy axis). The strength of MA is characterized by a magnetic field (anisotropy field $H_a$) which should be applied in the perpendicular direction (hard axis) to rotate the moment from the easy axis to the hard axis. The anisotropy energy $E_a$ is equal to the energy difference between the magnetic moment directed along the hard and the easy axis.

The key prerequisites of MA are the orbital moment, the spin-orbit (s-o) interaction and the anisotropic interactions of a magnetic ion with neighboring ions [1,2]. The s-o interaction, as a relativistic effect, becomes stronger in heavier ions. Consequently, the MA dominates the magnetism in *f*-electron materials.

The crystalline-electric-field (CEF) interaction is a mechanism of the MA in lanthanide compounds with well-localized 4*f*-electron magnetic moments [3–5]. The electron orbitals adopt orientations in the crystal lattice that minimize the energy of the CEF interaction. In contrast to the 4*f*-orbitals, deeply buried in the core electron density of lanthanide ions, the U 5*f*-electron wave functions are spatially extended. Consequently, they overlap with the 5*f*-orbitals of the neighbor U ions (5*f*-5*f* overlap) and with the orbitals of the valence electrons of other ligands which leads to hybridization of the 5*f*-electron states with the ligand valence-electron states (5*f*-ligand hybridization) [6].

The 5*f*-5*f* overlap allows for direct U-U exchange interaction (EI) whereas the 5*f*-ligand hybridization mediates the indirect 5*f*-ligand-5*f* exchange. These mechanisms cause that the 5*f* wave functions may lose their atomic character, the 5*f*-electrons become delocalized and the 5*f*-moments naturally reduced. Despite the 5*f*-electron delocalization, the strong s-o coupling induces a predominantly orbital magnetic moment antiparallel to the spin moment in the spin-polarized 5*f*-electron energy bands [7]. This gives rise to strong MA even in U intermetallics with strongly delocalized 5*f* electrons. The itinerant 5*f*-electron ferromagnet $UNi_2$ with a U magnetic moment of only few hundredths of a $\mu_B$, which exhibits a very strong MA with $H_a \gg 35$ T at 4.2 K, serves as an excellent example [8–11].

A much stronger MA than observed for lanthanide ions in an analogous crystal environment is inherent to U magnetism. The typical values of $H_a$ of most U intermetallic compounds so far studied are of the order of hundreds of teslas [12]. The interaction of the spatially extended U 5*f*-orbitals with ligands implies an essentially different mechanism of the MA, based on a two-ion (U-U) interaction. A relatively simple model of the two-ion interaction has been formulated by Cooper et al. [13,14] based on the Coqblin-Schrieffer approach to the mixing of ionic f-states and conduction-electron states [15]. The theory has been further extended so that each partially delocalized *f*-electron ion is coupled by the hybridization to the band electron sea leading to a hybridization-mediated anisotropic two-ion interaction causing MA [16].

When cooled in zero magnetic field, the bulk of a ferromagnetic (FM) material decomposes into mutually antiparallel FM domains (with the exception of flux closure domains) in order to minimize the magnetostatic energy. Neighboring domains are separated by domain walls of a thickness determined by the balance between $E_a$ and the exchange-interaction (EI) energy ($E_{ex}$). The EI favors a slow rotation of the magnetization while the MA prefers a sudden magnetization reversal. In the 3*d*-electron ferromagnets with a typically weak MA and strong EI ($E_{ex} \gg E_a$) the domain-wall thickness may be even hundreds of distance between nearest magnetic ions. On the other hand, the strong MA ($E_a \gg E_{ex}$) in $f$-electron ferromagnets may cause reduction of the domain wall width down to a few interatomic distances [17,18].

A number of U$TX$ compounds with transition metals ($T$) and $p$-electron metals ($X$) crystallize in the hexagonal ZrNiAl-type structure [12] (space group P-62m, No. 189), shown in FIG. 1a). The structure is characterized by stacking of U-$T$1 and $T$2-$X$ basal-plane layers along the $c$-axis, where $T$1 and $T$2 are two different positions for the transition metal $T$. The U ions in the U-$T$1 basal plane are arranged in a slightly distorted kagome lattice (Fig. 1b)) in which each U ion has four U nearest neighbors at a shortest U-U distance. This arrangement leads to considerable 5$f$-5$f$ overlap and a 5$f$(U)-d($T$) ligand hybridization which leads to a compression of the 5$f$-electron density towards the basal plane. This phenomenon assisted by the orbital-moment polarization results in a strong uniaxial MA with the easy axis perpendicular to the basal plane, i.e. along the $c$-axis. This suits the more general scenario of the easy axis oriented perpendicular to the nearest U-U bonding links in the lattice [19]. Huge $H_a$ values of the order of hundreds of teslas have been estimated for the isostructural U$TX$ ferromagnets [20–22] from the intercepts of the easy- and hard-magnetization curves extrapolated to high magnetic fields. The rather low values of the Curie temperature ($T_c$) of the order of tens of kelvins [12] indicate that $E_a \gg E_{ex}$, which leads to the occurrence of very narrow domain walls.

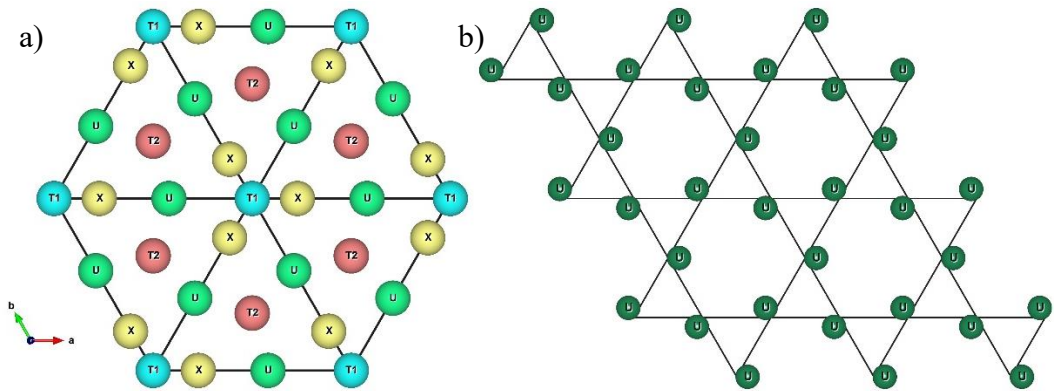

FIG. 1. a) Hexagonal ZrNiAl- type structure of U$TX$ compounds. Two different position of the transition metal T are shown as T1 and T2. The c-direction is perpendicular to the plane, as a result U-$T$1 and $T$2-$X$ planes are shown together. b) Schematic representation of magnetic U ions creating a distorted kagome lattice with the non-distorted kagome lattice for comparison. The c-direction is perpendicular to the plane.

The direct imaging and investigation of domain structures in U compounds have been done on polycrystalline samples at room temperature (RT) in materials with a high concentration of 3$d$ transition metals [23,24] with magnetic moments which are ferromagnetically coupled by strong EI resulting in high $T_C$-values ($\gg$ RT). The only report on the single crystalline material containing uranium has been recently published [25] on the UMn$_2$Ge$_2$. Although this material has comparable [26] anisotropy (roughly ¼) in comparison to the compound under study [27], the formation of magnetic domains is determined by the ordering of Mn moment at higher (above RT) temperatures.

In this paper, we present the direct low-temperature investigation of a domain structure in a U 5$f$-electron moment ferromagnet with $T_C \ll$ RT. UCoGa, a member of the family of U$TX$ compounds with ZrNiAl-type of structure, has been selected as a typical U ferromagnet with huge uniaxial MA [20,27–30] having the easy axis along the c-axis of the hexagonal structure. First, the UCoGa single crystal has been characterized by magnetization measurements in magnetic fields applied along the main crystallographic axes, *a* and *c*. The domain-structure imaging has been accomplished by using a low-temperature atomic force microscope - attoAFM/MFM Ixs (attocube). Analysis of the available data has led to the conclusion that the huge uniaxial MA, together with the relatively weak EI in UCoGa and possibly in most other U ferromagnets, causes extremely narrow magnetic domain walls with width equal to the distance between the nearest magnetic U neighbors with antiparallel magnetic moments.

**Experimental**

The growth and thermal treatment of the UCoGa single crystal are described in [29]. An approximately 1 mm thick disk has been cut perpendicular to the crystallographic *c*-axis from an annealed ingot. The flat surface was polished with diamond paste of 1 µm grade. The surface was then washed in acetone and isopropanol. The surface topography, measured by atomic force microscopy (AFM) in tapping mode at 50 K, is presented in FIG. 2. From the same area, MFM images were acquired. The concentration of scratches and dirt particles was below the acceptable limit for measurements in the lift-mode. These objects served as markers on the surface.

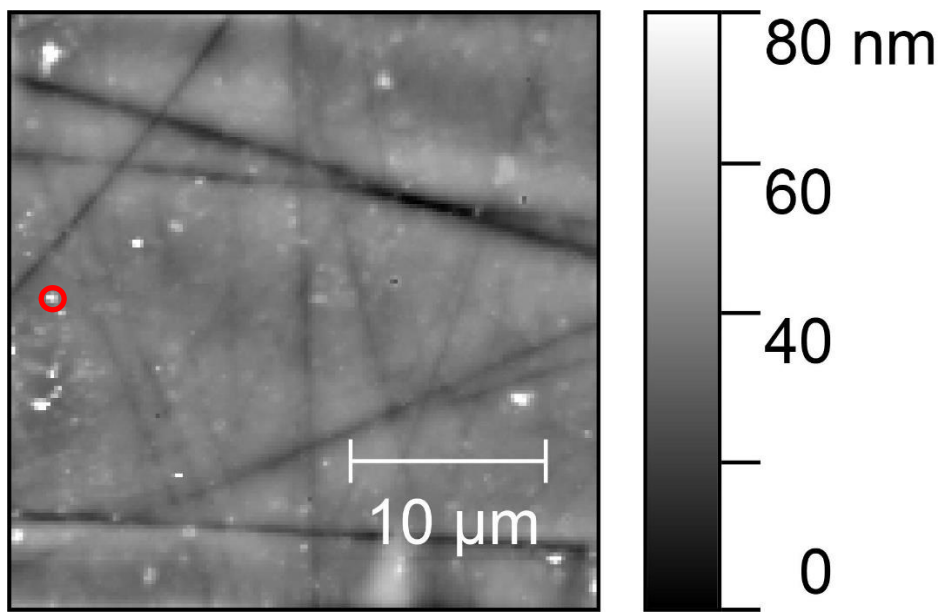

FIG. 2. Topography image (AFM) of the sample surface at 50 K. The MFM data were collected from the same area. The dirt particle in the red circle is used as a position marker in all images.

Magnetization measurements were performed in a SQUID magnetometer MPMS-7-XL and a PPMS-14T (both Quantum Design) on the same sample as used for the MFM studies. The AFM and MFM measurements were performed using a low-temperature atomic force microscope - attoAFM/MFM Ixs inserted in a PPMS-14T. PPP-MFMR probes by NANOSENSORS with hard-magnetic coating (coercivity of about 300 Oe) were used. The MFM measurements were done in the constant-height mode at a lift-height of 80-100 nm. The frequency modulation mode was applied in which the cantilever oscillates at its resonance frequency and the change of resonance frequency d*f* represents the magnetic-force contrast. The residual field of the PPMS-14T superconducting magnet was minimized prior to the MFM measurement. This allowed to measure zero-field-cooled images with (almost) equal distribution of up- and down-magnetized domains.

The width of the surface domains was determined by the stereological method in which it is calculated from the number of intersections of an arbitrary test line with domain walls in the MFM image ($W_s = 2l/\pi n$, where $l$ is the total test-line length and $n$ the number of intersections) [31], 40 lines per MFM image were evaluated. The error was calculated as three standard deviations of a normal distribution.

**Results**

The temperature dependence of the magnetization of UCoGa in a magnetic field of 10 mT applied along the $c$-axis is shown in FIG. 3. $T_C = 45$ K has been estimated as the temperature of the middle of the sharp decrease of the $M$ vs $T$ dependence.

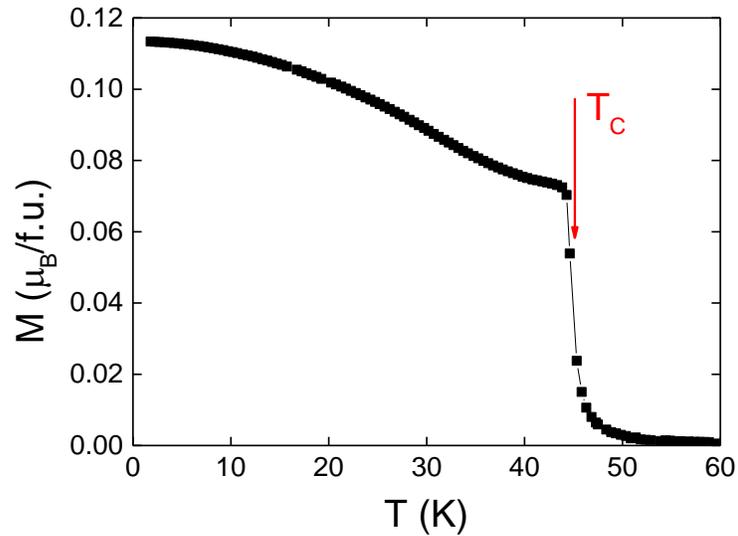

FIG. 3. Temperature dependence of the magnetization in an external magnetic field of 10 mT applied along the $c$-axis.

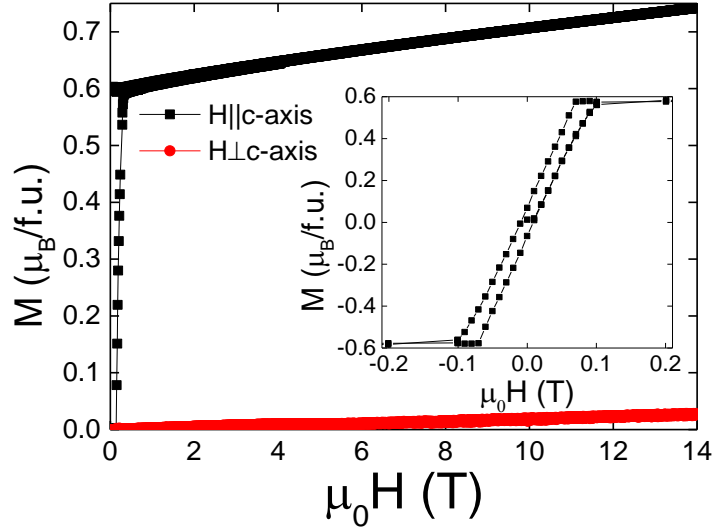

FIG. 4. Magnetization isotherms measured at 2 K in fields applied parallel and perpendicular to the *c*-axis. The inset shows the low-field zoom of a magnetic isotherm (showing the details of the hysteresis) measured at 20 K in a field applied along the *c*-axis.

The magnetization isotherms measured at 2 K in fields applied parallel and perpendicular to the *c*-axis shown in FIG. 4 clearly demonstrate the huge uniaxial MA of UCoGa. The *c*-axis magnetization which is equal to 0.61 $\mu_B$/f.u. at 1 T slowly saturates with increasing the field and reaches the value of 0.75 $\mu_B$/f.u. The *a*-axis magnetization shows a very weak linear (paramagnetic) response to the magnetic field and reaches only a value of 0.03 $\mu_B$/f.u. in 14 T. When the *a*-axis *M* vs. *H* straight line is extrapolated to high magnetic fields it reaches 0.75 $\mu_B$/f.u. at $\mu_0 H$ = 350 T. This number can be tentatively considered as the minimum estimated $H_a$ value. It is considerably larger than $H_a \geq 150$ T derived from magnetization data collected at 4.2 K in fields up to 35 T [20].

The inset of Fig. 4 shows a low-field detail of the hysteresis loop measured in field along the *c*-axis field at 20 K. At this temperature, the coercive field is about 0.3 T. The coercivity collapses upon heating to temperatures above 26 K. This finding is in agreement with the results reported in Refs [20] and [29].

In the case of a ferromagnet with strong uniaxial anisotropy, a domain pattern created by domain branching can be observed at a surface perpendicular to the easy axis [31]. The evolution of magnetic domains during the magnetization process in UCoGa, studied at 20 K by means of MFM is presented in Fig. 5. Starting with the zero-field-cooled (ZFC) image (5a), one can see a labyrinth-like domain structure. At 0.01 T (5b), a very similar domain pattern is observed. The "dark" domains (magnetized in the same direction as the probe) expand with further increasing magnetic field. A partial magnetization is observed, e.g. at 0.05 T (5c). For $\mu_0 H$ = 0.1 and 0.3 T (5d and 5e, respectively), the domains disappear which is consistent with the saturation of magnetization in the inset of Fig. 4. New domains appear for $\mu_0 H < 0.05$ T (5f and 5h) when the sample becomes demagnetized from the saturated state when sweeping the field back down (see inset of Fig. 4). These domains have a different shape than those observed in the ZFC state (Fig. 5a). Nevertheless, some of them are pinned to the same surface point where a lattice defect can be expected. The lattice defects, which serve as centers of pinning of the magnetic domain walls, are usually unaffected by the applied magnetic field and, therefore, they are fixed in the lattice in the magnetization/demagnetization process. An analogous magnetization process was observed in a negative applied magnetic field.

Weak parallel lines, forming a diamond like pattern (indicated by the green dashed lines), persist in the magnetic contrast (5d-f) and cannot be removed by an applied magnetic field up to 14 T (not shown). The pattern lines cross with angles of 120° and are probably related to the structure defects mentioned above. In the partially magnetized state, some of the magnetic domain walls are pinned to these lines.

The MFM images in Fig. 6 show the evolution of magnetic domains upon ZFC cooling. The magnetic domains are formed just below $T_C$. The d$f$ contrast increases upon cooling as expected for increasing magnetic moments of the domains. A gradually developing domain-branching pattern is observed upon cooling from 40 K to lower temperatures. The shape of the domains at the surface changes only slightly, with decreasing temperature emphasizing the morphological details. The magnetic domains resemble the ZFC domains at 20 K (5a), only the contrast is inverted due to the opposite magnetization of the probe (due to the magnetic history of the previous measurements).

The width of the surface domains $W_s$ decreases with temperature decrease as can be seen in Fig. 7. This decrease reflects the narrowing of the magnetic domains inside the sample with increasing magnetic moment.

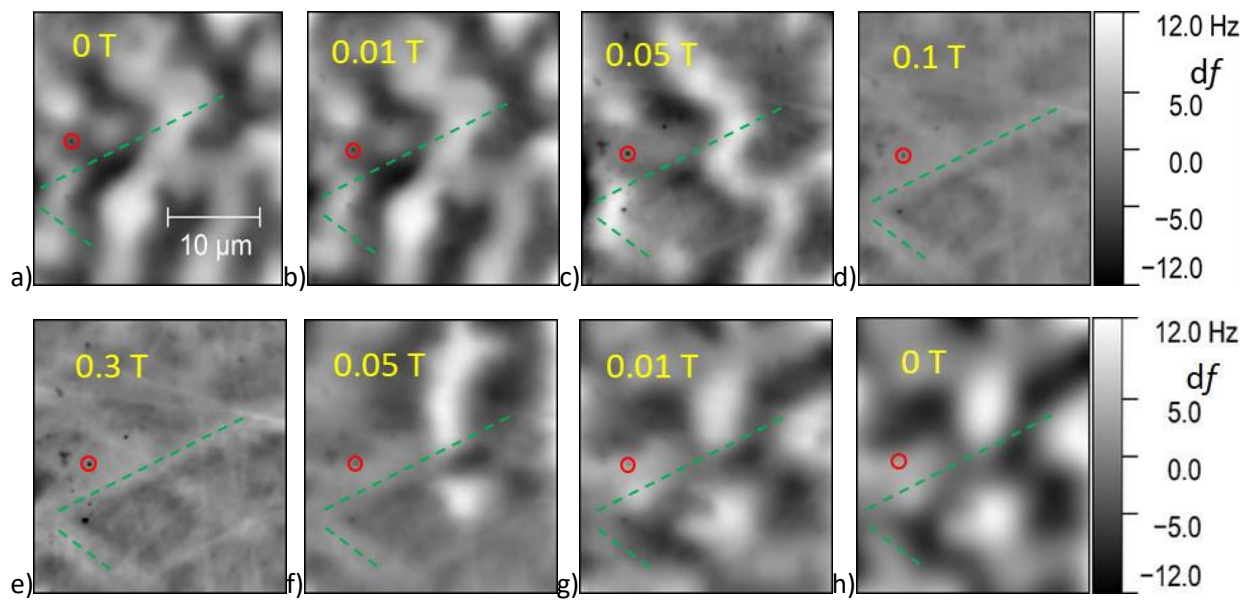

FIG. 5. MFM scans at 20 K in increasing and decreasing magnetic field. All scans from a) to h) were made during increasing, followed by decreasing, of the applied magnetic field. The scan in a) was done at zero field after cooling from high temperatures. The scans in a) 0 T (ZFC state), b) 0.01 T, c) 0.05 T, d) 0.1 T and e) 0.3 T show the process in increasing magnetic field. The process in the decreasing field is shown in the scans f) 0.05 T, g) 0.01 T and h) 0 T. The MFM probe was magnetized in a positive magnetic field.

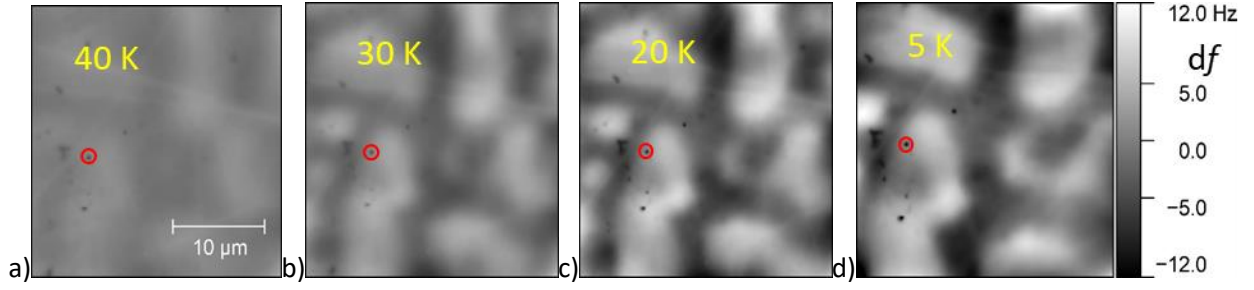

FIG. 6. Magnetic domains of UCoGa at a) 40 K, b) 30 K, c) 20 K and c) 5 K.

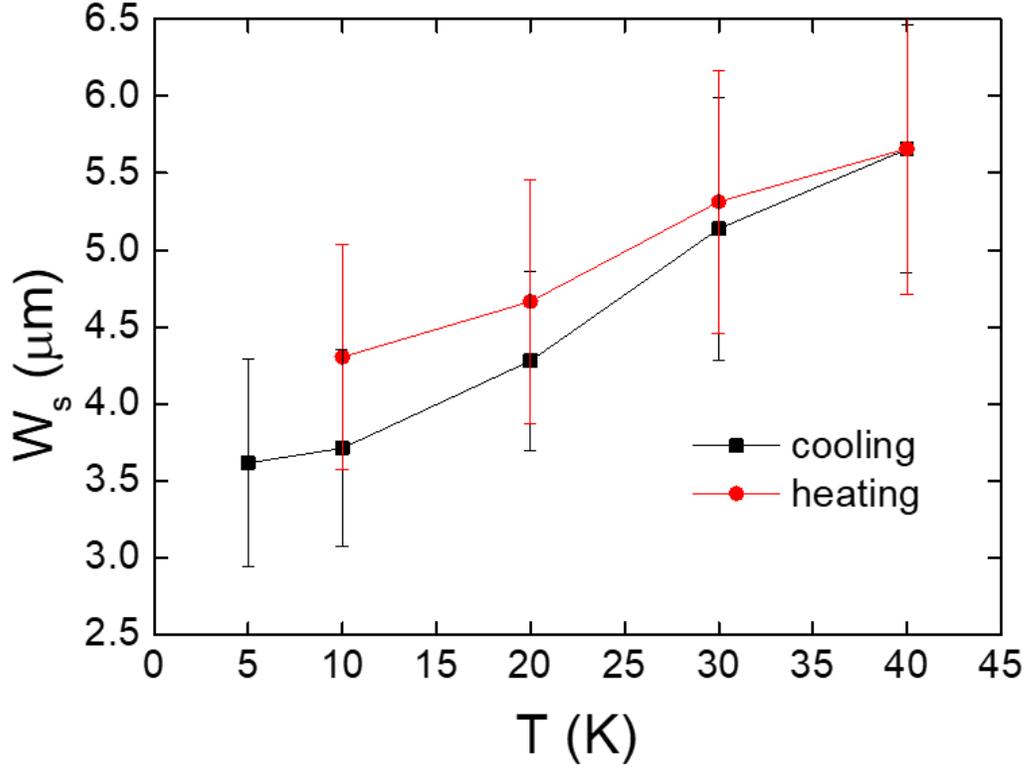

FIG. 7. Evolution of the domain width with temperature upon heating and cooling.

The domain-wall energy $\gamma_W$ in a ferromagnet with a strong uniaxial anisotropy can be calculated by using the equation

$$\gamma_W = \frac{W_s m_s^2}{49\mu_0}, \qquad (1)$$

where $m_s$ is the spontaneous magnetization and $\mu_0$ the permeability of vacuum [31–33]. A $\gamma_W$ value of 1.1 mJ/m$^2$ is obtained with the $W_s$ and $m_s$ values, determined at the lowest temperature of measurement (5 K) where the thermal effects are minimized.

Most FM materials are characterized by Bloch domain walls. However, a simple calculation shows that this is not the case for UCoGa. The width of a Bloch domain wall in UCoGa can be calculated from the equation

$$\delta = \frac{\gamma_W}{K_1}, \tag{2}$$

where $K_1$ is the anisotropy constant [34]. Using the value $K_1$ = 88 MJ/m$^3$ reported by Prokes et al. [27] and the above given domain-wall energy of 1.1 mJ/m$^2$, a Bloch domain wall thickness $\delta$ = 12.5 pm is obtained. This value is about 30 times smaller than the distance between two nearest magnetic U ions (350 pm [35]) in UCoGa. The failure of the Bloch-domain-wall model is not surprising. In this model, the magnetic moment rotates between neighbouring magnetic ions by an angle of about 1° in case of weak MA and/or strong EI, which is typical for the 3d-electron ferromagnets. In contrast, in the case of UCoGa and most U 5$f$-electron ferromagnets, the MA is very strong while EI is rather weak.

If we consider a large angle of rotation between neighbouring magnetic moments, 180° in particular, the domain wall is a slab (oriented along the c-axis) consisting of pairs of nearest U neighbors (within the basal plane) having antiparallel magnetic moments aligned along the $c$-axis. The domain wall thickness is than equal to the distance between the neighboring magnetic ions. In such a case, the contribution of the uniaxial MA to the domain-wall energy is zero because both directions of magnetic moments in the wall are energetically equal. The energy needed to create a domain is then simply equal to the EI energy between the nearest magnetic moments.

The results of our recent study of the critical exponents suggest that UCoGa belongs to the universality class of the 2D Ising system with long-range magnetic order [36]. The 2D Ising character has recently been reported also for the isostructural compound URhAl [37]. The crystal structure of UCoGa is characterized by a distorted kagome lattice of U ions as schematically shown in Fig. 1b). The EI energy can be derived from $T_C$ using the relation for the 2D Ising ferromagnet kagome lattice [38]. The smallest element of a domain wall in UCoGa is a trio of U ions forming an equilateral triangle with two U ions with parallel moments and one U ion with an antiparallel moment. The EI energy such a trio of moments is equal to

$$J = k_B T_C \frac{\ln(3+2\sqrt{3})}{2}, \tag{3}$$

and when divided by the area of the smallest element of a domain wall, it leads to $\gamma_W$ = 1.4 mJ/m$^2$. This value is in a reasonable agreement with the value of 1.1 mJ/m$^2$ determined above from the surface domain width. The difference between the two $\gamma_W$ values may be also due to neglecting some contributions to the EI.

For simplicity, only the short-range direct FM 5$f$-5$f$ interaction has been considered in our scenario. The role of hybridization of the magnetic U 5$f$-electron states with the Co 3$d$ states has been neglected. Small magnetic moments on Co sites may be expected as a natural consequence of the 5$f$-3$d$ ligand hybridization. No relevant microscopic study in UCoGa has been done to prove this expectation. Nevertheless, the detailed polarized-neutron-diffraction experiments on the isostructural U*TX* compounds, UCoAl in the metamagnetic state [39] and the ferromagnets URhAl [40] and UPtAl [41] below $T_C$ have unambiguously revealed the existence of small magnetic moments (parallel to the U moments) at the *T* ions on the inequivalent crystallographic positions in the U-*T*1 and *T*2-*X* basal-plane layers, respectively. These findings corroborate that the expectation of small induced Co moments in UCoGa is justified. The responsible 5*f*-3*d* hybridization also mediates additional indirect FM U-Co-U components to the hierarchy of exchange interactions in UCoGa. Relevant neutron and X-ray magnetic-scattering experiments would help to confirm this scenario.

**Conclusions**

UCoGa exhibits a surface-domain structure typical for a strongly anisotropic uniaxial ferromagnet which is consistent with the huge uniaxial MA observed in this compound. The MFM images show the appearance/disappearance of magnetic domains during the magnetization/demagnetization processes with increasing/decreasing applied magnetic field. In the partially magnetized state, the magnetic domain walls are pinned to lattice defects which have been identified in the AFM and MFM images. The evolution of magnetic domains during cooling the UCoGa crystal below $T_C$ has been also demonstrated by MFM images. Analysis of available data leads to the conclusion that the high energy of the uniaxial magnetocrystalline energy assisted by the relatively low energy of ferromagnetic exchange interaction in UCoGa causes that the domain walls are extremely narrow, equal to the distance between the nearest magnetic U neighbor ions with antiparallel magnetic moments in the U-*T*1 basal plane of the hexagonal structure. Analogously, the same unique domain-wall properties are expected for other uniaxial U ferromagnets which are characterised by high anisotropy energy in combination with moderate values of the exchange energy. More MFM and magneto-optic Kerr effect studies of magnetic domains in these materials at low temperatures are desired to support the conclusions.

**Acknowledgements**

The authors would like to thank Prof. F. R. de Boer for critical reading and correcting the manuscript. This research was supported by the Czech Science Foundation, grant No. 16-06422S. Experiments were performed in MGML (www.mgml.eu), which is supported within the program of Czech Research Infrastructures (project no. LM2018096).